%% file: BSSBookch10.tex
\documentclass[graybox, envcountchap]{svmult}

\usepackage{mathptmx}        
\usepackage{helvet}          
\usepackage{courier}         

\usepackage{makeidx}         
\usepackage{graphicx}        
\usepackage{multicol}        
\usepackage[bottom]{footmisc}

\makeindex             

\input{common2}

\begin{document}
\pagestyle{empty}
\frontmatter

\include{dedic}
\include{foreword}
\include{preface}

\mainmatter

\include{mcmillan}
\backmatter
\printindex


\end{document}

%% file: common2.tex
\usepackage{amsmath}
\usepackage{amssymb}
\usepackage{lscape}
\usepackage{multirow}

\def\gapp{\ifmmode\stackrel{>}{_{\sim}}\else$\stackrel{<}{_{\sim}}$\fi}
\def\gsim{\lower.5ex\hbox{\gtsima}}
\def\gtsima{$\; \buildrel > \over \sim \;$}
\def\lapp{\ifmmode\stackrel{<}{_{\sim}}\else$\stackrel{<}{_{\sim}}$\fi}
\def\lsim{\lower.5ex\hbox{\ltsima}}
\def\ltsima{$\; \buildrel < \over \sim \;$}

\newcommand\apgt{\ {\raise-.5ex\hbox{$\buildrel>\over\sim$}}\ }
\newcommand\aplt{\ {\raise-.5ex\hbox{$\buildrel<\over\sim$}}\ }

\newcommand{\half}{{\ensuremath{\textstyle\frac12}}}
\newcommand{\kin}{{\ensuremath K}}

\newcommand{\mbar}{{\ensuremath \langle m\rangle}}
\newcommand{\msun}{{\ensuremath {\rm M}_\odot}}
\newcommand{\pot}{{\ensuremath U}}
\newcommand{\quarter}{{\ensuremath{\textstyle\frac14}}}
\newcommand{\rh}{{\ensuremath R_h}}
\newcommand{\rj}{{\ensuremath R_J}}
\newcommand{\rsun}{{\ensuremath R_\odot}}
\newcommand{\rvir}{{\ensuremath R_{vir}}}
\newcommand{\tcc}{{\ensuremath t_{cc}}}
\newcommand{\tdis}{{\ensuremath t_{dis}}}
\newcommand{\tdyn}{{\ensuremath t_{dyn}}}
\newcommand{\tenc}{{\ensuremath t_{enc}}}
\newcommand{\third}{{\ensuremath{\textstyle\frac13}}}
\newcommand{\threehalf}{{\ensuremath{\textstyle\frac32}}}
\newcommand{\trh}{{\ensuremath t_{rh}}}
\newcommand{\trl}{{\ensuremath t_{rl}}}
\newcommand{\tr}{{\ensuremath t_r}}
\newcommand{\tseg}{{\ensuremath t_{seg}}}
\newcommand{\va}{{\bf a}}
\newcommand{\vdisp}{{\ensuremath \langle v^2\rangle}}
\newcommand{\vesc}{{\ensuremath v_{esc}}}
\newcommand{\vinf}{{\ensuremath v_\infty}}
\newcommand{\vrms}{{\ensuremath \vdisp^{1/2}}}

\newcommand{\vx}{{\bf x}}

%% file: mcmillan.tex
\setcounter{chapter}{9}

\title{Dynamical Processes in Globular Clusters}
\titlerunning{Globular Cluster Dynamics}
\author{Stephen L. W. McMillan}
\institute{Stephen McMillan \at Department of Physics, Drexel
  University, Philadelphia, PA 19104, USA,\\
  \email{steve@physics.drexel.edu}}

\maketitle
\label{Chapter:McMillan}

\abstract*{Globular clusters are among the most congested stellar
  systems in the Universe.  Internal dynamical evolution drives them
  toward states of high central density, while simultaneously
  concentrating the most massive stars and binary systems in their
  cores.  As a result, these clusters are expected to be sites of
  frequent close encounters and physical collisions between stars and
  binaries, making them efficient factories for the production of
  interesting and observable astrophysical exotica.  I describe some
  elements of the competition among stellar dynamics, stellar
  evolution, and other processes that control globular cluster
  dynamics, with particular emphasis on pathways that may lead to the
  formation of blue stragglers.}

\section{Introduction}
\label{McMillan:Sec:introduction}

Globular clusters\index{globular cluster} have long been regarded as near-perfect laboratories
for studies of stellar physics\index{stellar physics} and stellar dynamics\index{stellar dynamics}.  Some reasons
(and complications) are:
\begin{itemize}
\item they are isolated in space (but not all clusters are found in
  galactic halos  ---  many disk and bulge clusters are known, and
  dynamical friction has probably transported many clusters into the
  Galactic Centre)
\item they contain coeval stars (but many clusters are now known to
  contain multiple stellar populations\index{multiple stellar population} indicating several distinct
  phases of star formation)
\item they contain virtually no gas or dust (today, that is --- at early
  times, gas dynamical processes dominated their evolution)
\item they are nearly spherical (although several are measurably
  flattened by rotation and/or tidal effects)
\end{itemize}

These systems thus represent a relatively --- although not
perfectly --- ``clean'' realisation of the classical $N$-body problem\index{N-body problem}
\begin{equation}
    \va_i ~\equiv~ \ddot{\vx}_i ~=~ \sum_{j\ne i}^N\ Gm_j
			\frac{\vx_j-\vx_i}{\left|\vx_j-\vx_i\right|^3},
			~~~i = 1,\ldots,N.
\label{McMillan:Eq:nbody}
\end{equation}
We begin our study of cluster dynamics by ignoring complicating
factors such as gas dynamics, stellar evolution, mass loss, etc., and
focus on the pure $N$-body problem, much as it might have been
described by Newton.  We define timescales and other units, discuss
the fundamental dynamical processes driving cluster evolution, and
present some basic terminology relevant to cluster dynamics.

\section{Virial Equilibrium}
\label{McMillan:Sec:dynamics}

Star clusters have no static equilibrium configuration similar to that
found in a fluid system such as a star.  Stars are in constant motion.
However, in dynamical equilibrium, at any given location in the
cluster there are as many stars moving inward as moving outward --- that
is, there is no net radial stellar flux.

\subsection{The Virial Theorem}\index{virial theorem}
\label{McMillan:Subsec:virial}

A convenient global restatement of dynamical equilibrium\index{dynamical equilibrium} involves the
``radial moment of inertia'' of the system
\begin{equation}
    I = \sum_{i=1}^N\,m_i\,r_i^2,
\end{equation}
where $r_i=|{\bf x}_i|$.  Differentiating, we find
\begin{equation}
    \ddot{I} = 2\sum_{i=1}^N\, m_i\left(v_i^2 + \vx_i\cdot\va_i\right).
\end{equation}
Setting $\ddot{I}=0$ as our definition of dynamical equilibrium, we
have
\begin{equation}
    \sum_{i=1}^N\, m_i\,v_i^2 + \sum_{i=1}^N\,m_i\,\vx_i\cdot\va_i ~=~ 0.
\end{equation}
The first term is simply twice the total kinetic energy of the system,
$2\kin$.  The second is easily shown to be
\begin{equation}
    {\pot} = -\sum_{i=1}^N\,\sum_{j>i}^N\ 
			\frac{Gm_im_j}{\left|\vx_j-\vx_i\right|},
\end{equation}
the total potential energy\index{potential energy} of the system.  Thus we obtain the (scalar)
Virial Theorem
\begin{equation}
    2\kin + {\pot} = 0.
\label{McMillan:Eq:virial}
\end{equation}
If this relation holds the cluster is said to be in virial
equilibrium.  Since the total energy is $E = \kin + \pot ~(< 0)$, in
virial equilibrium we have
\begin{equation}
    \kin ~=~ -E,~~~~\pot ~=~ 2E.
\label{McMillan:Eq:specificheat}
\end{equation}

\subsection{Length and Time Scales}
\label{McMillan:Subsec:scales}

We can define some characteristic physical scales for a system in
virial equilibrium (Eq.~\ref{McMillan:Eq:virial}).  For a cluster of
total mass $M$, the virial radius, $\rvir$, is defined as
\begin{equation}
    \rvir ~\equiv~ -\frac{GM^2}{2\,\pot} ~=~ -\frac{GM^2}{4E}.
\end{equation}
It defines a characteristic length scale for the cluster.  It is
typically comparable to the cluster half-mass radius\index{half-mass radius}, $\rh$, the
radius of the sphere centred on the cluster enclosing half of the
cluster's total mass.  The two radii are often used interchangeably,
although they are distinct physical quantities.  Spitzer \cite{1987degc.book.....S} notes
that $\rvir\approx0.8\rh$ for a broad range of common cluster models.

The cluster dynamical time\index{dynamical timescale} (or ``crossing time'')\index{crossing time}, $\tdyn$, is
\begin{eqnarray}
    \tdyn &\equiv& \left(\frac{GM}{\rvir^3}\right)^{-1/2}
		~=~ \frac{GM^{5/2}}{(-4E)^{3/2}}\label{McMillan:Eq:tdyn}\\
\noalign{\medskip}
	  &=& 0.47~\mbox{Myr}
		   \left(\frac{M}{10^6\,\msun}\right)^{-1/2}
		   \left(\frac{\rvir}{10\,\rm pc}\right)^{3/2}.\nonumber
\end{eqnarray}
The second forms of this and the previous expression conveniently
define $\rvir$ and $\tdyn$ in terms of conserved quantities.  The
dynamical time\index{dynamical time} is the characteristic orbital or free-fall time\index{free-fall time} of a
cluster.  It is also the timescale on which an initially
non-equilibrium cluster will establish virial equilibrium.  Since
$\tdyn$ is generally short compared to all other dynamical timescales
of interest, we assume virial equilibrium in all that follows.

Finally, the cluster-wide velocity dispersion $\vdisp$ is
\begin{eqnarray}
    \vdisp &=& \frac{2\kin}{M} ~=~
	    \frac{GM}{2\rvir}\label{McMillan:Eq:vdisp_rvir}\\
	   &=& (14.7~\mbox{km/s})^2 
		   \left(\frac{M}{10^6\,\msun}\right)
		   \left(\frac{\rvir}{10\,\rm pc}\right)^{-1}.\nonumber
\end{eqnarray}
Dynamicists commonly write the total kinetic energy $\kin$ in terms of
the ``thermodynamic'' quantity $kT$, defined by
\begin{equation}
    \kin ~=~ \threehalf NkT
\end{equation}
so
\begin{equation}
    kT ~=~ \third\mbar\vdisp ~=~ -\frac{2E}{3N}
			 ~~~~ \mbox{in virial equilibrium},
\label{McMillan:Eq:kT}
\end{equation}
where $\mbar=M/N$ is the mean stellar mass.

Since gravity has no preferred scale, it is convenient to work in a
system of dimensionless units such that all bulk cluster properties
are of order unity.  A system in widespread use, described in \cite{HM86}, 
has $G=1, M=1$, and $E=-\quarter$, so $\rvir=1,
\tdyn=1$, and $\vdisp=\half$.

\section{Relaxation}
\label{McMillan:Sec:relaxation}

The long-term evolution of a star in hydrostatic equilibrium\index{hydrostatic equilibrium} is driven
by thermal and nuclear processes that transfer energy throughout the
star and generate energy in the core\index{stellar core}.  In a star cluster, thermal
evolution\index{thermal
evolution} is driven by two-body relaxation\index{two-body relaxation}, while energy may be
generated by a number of mechanisms, as discussed below.

\begin{figure}[h]
\includegraphics[height=50mm]{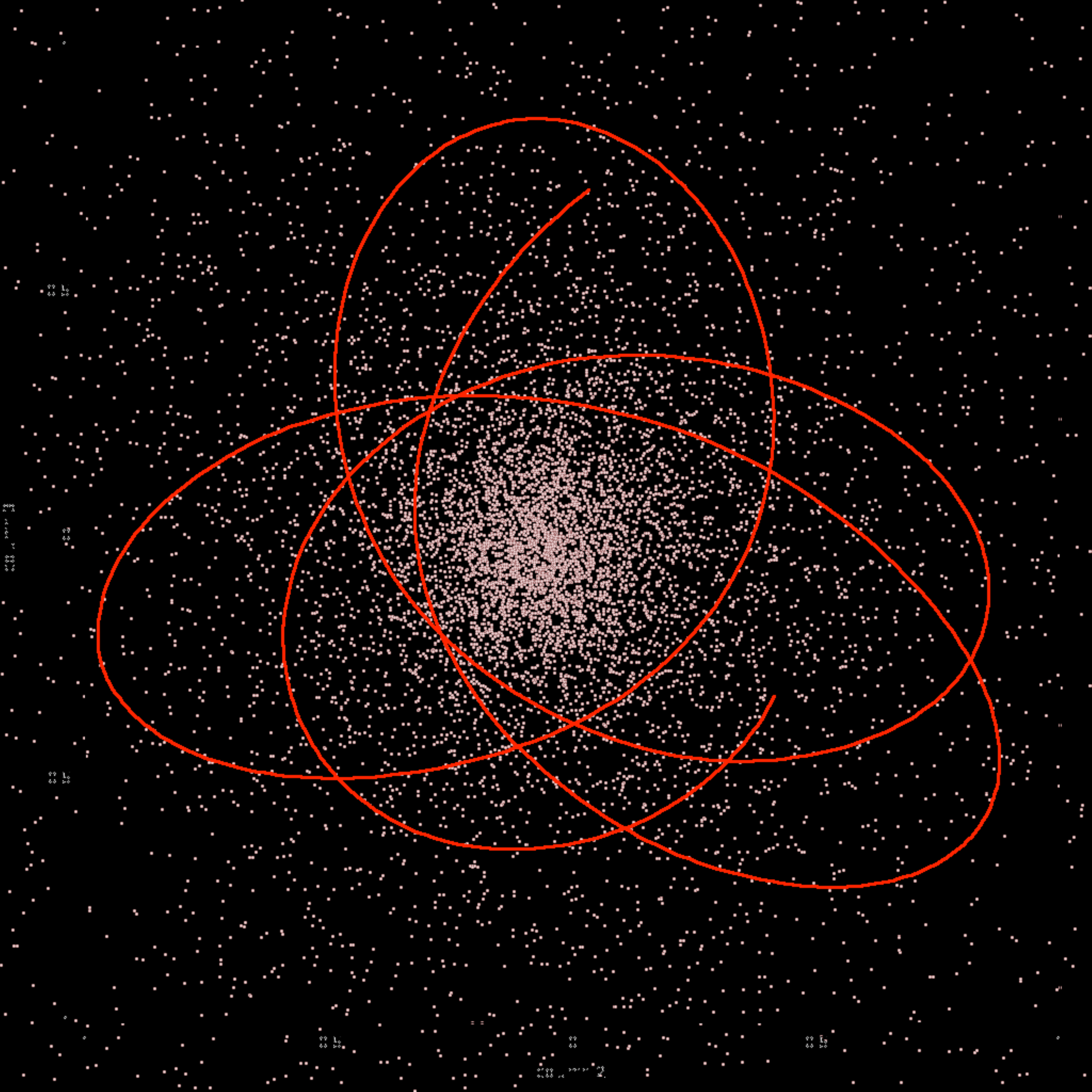}
~\includegraphics[height=50mm]{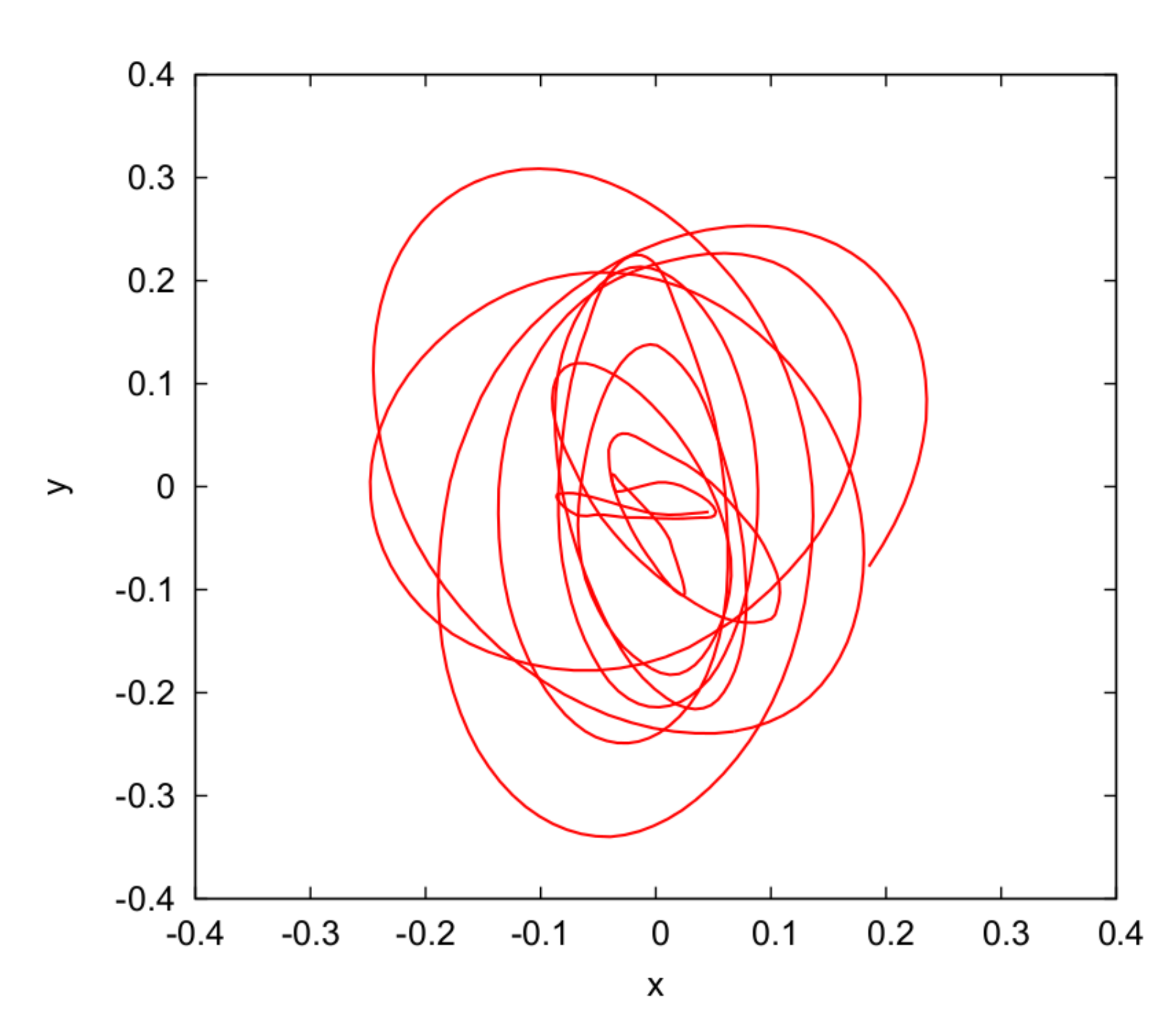}
\caption{(a) A typical smooth orbit in a 10,000-body system.  The
  scale of this figure is $\pm1$ $N$-body units.  (b) An orbit closer
  to the centre shows generally smooth behaviour, but also has a few
  sharp ``kinks'' associated with close encounters in the dense core.}
\label{McMillan:Fig:orbits}
\end{figure}

To a first approximation (Fig.~\ref{McMillan:Fig:orbits}a), stars
orbiting in a cluster move on relatively smooth orbits determined by
the bulk mean-field gravitational potential\index{gravitational potential} of the system as a whole.
However, stars occasionally experience close encounters with one
another, changing their orbital parameters and transferring energy
from one to the other (see Fig.~\ref{McMillan:Fig:orbits}b).
This thermalising process allows energy to flow around the stellar
system.

\subsection{Two-body Scattering}
\label{McMillan:Subsec:scattering}

Our basic approximation here is the assumption that encounters between
stars can be treated as isolated two-body scattering\index{two-body scattering} events.  This is
permissible because the scale of two-body encounters is generally much
less than the scale of the system, so we can talk sensibly about
stellar velocities at ``infinity'' in a scattering calculation without
having to worry about the large-scale motion of stars around the
cluster.

\begin{figure}[h]
\includegraphics[height=55mm]{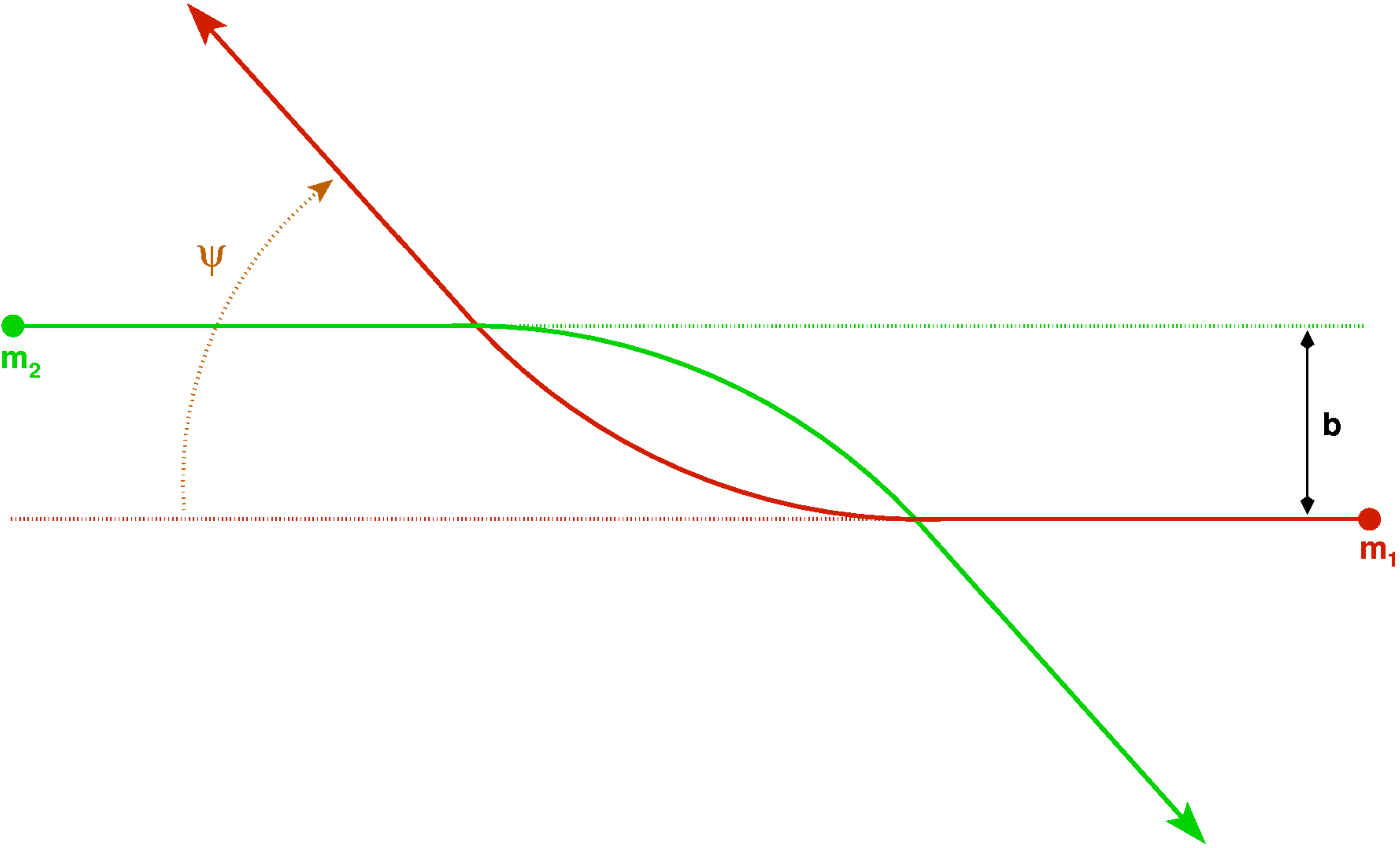}
\caption{Two stars, of masses $M_1$ and $M_2$, approach one another
  with impact parameter $b$ (and relative velocity at infinity
  $v_\infty$), and are deflected by an angle $\psi$.}
\label{McMillan:Fig:scattering}
\end{figure}

Imagine two stars of masses $M_1$ and $M_2$ approaching one another on
unbound trajectories with relative velocity at infinity $\vinf$ and
impact parameter $b$ (Fig.~\ref{McMillan:Fig:scattering}).  The
solution for the relative orbit ${\bf r} = \vx_1-\vx_2$ is
$$
    r(1+e\cos\theta) = a(e^2-1),
$$
where (with $m=M_1+M_2$), $a = Gm/\vinf^2$ is the semi-major axis, and
$e = \sqrt{1 + \left(b\vinf/Gm\right)^2}$ is the eccentricity.  The
deflection angle is $\psi = \pi-2\theta_1$, where
\begin{equation}
    \tan\theta_1 = \frac{b\vinf^2}{Gm}\,.
\end{equation}
Thus the impact parameter corresponding to a $90^\circ$
scattering, $\psi = \pi/2$ or $\theta_1=\pi/4$, is
\begin{equation}
    b_{90} ~=~ \frac{Gm}{\vinf^2}\,.
\end{equation}
More generally, for encounters in a cluster, we have $m\sim2\mbar$ and
$\langle\vinf^2\rangle\sim2\vdisp$, and we may write
\begin{equation}
    b_{90} ~\sim~ \frac{G\mbar}{\vdisp}\,.
\label{McMillan:Eq:b90}
\end{equation}
For $\mbar\sim1\msun$ and $\vdisp^{1/2}\sim10$ km/s,
Eq.~(\ref{McMillan:Eq:b90}) gives $b_{90}\sim9$ AU.  Combining
Eqs.~(\ref{McMillan:Eq:vdisp_rvir}) and (\ref{McMillan:Eq:b90}), we
find $b_{90}\sim2\rvir/N$.

\subsection{Strong Encounters}
\label{McMillan:Subsec:closeencounters}

The strong encounter timescale, $t_s$, is the time needed for a
typical star to experience a $90^\circ$ scattering.  For a star of
mass $m_\ast$ moving with velocity $v$ through a uniform field of
identical stars with number density $n$, the cross section for a
strong encounter is
\begin{equation}
    \sigma ~=~ \pi b_{90}^2
	   ~=~ \frac{\pi G^2m_\ast^2}{v^2}\,.
\end{equation}
The timescale for a strong encounter is
\begin{equation}
    t_s ~=~ (n \sigma v)^{-1} ~=~ \frac{v^3}{\pi G^2 m_\ast^2 n}\,.
\end{equation}
Replacing $m_\ast$ by the mean stellar mass $\mbar$, $v^2$ by the
stellar velocity dispersion $\vdisp$, and writing $\mbar n = \rho$, we
obtain
\begin{equation}
    t_s ~=~ \frac{\vdisp^{3/2}}{\pi G^2\mbar\rho}\,.
\label{McMillan:Eq:ts}
\end{equation}
This is the relevant timescale for discussions of interactions
involving close binaries (see \S\ref{McMillan:Subsubsec:binaries}
below).

\subsection{Distant Encounters}
\label{McMillan:Subsec:distantencounters}

The cross section\index{cross section} for wide encounters, with smaller deflections
$\psi\ll1$, is much larger than that for a $90^\circ$ scattering, but
to estimate the cumulative effect of many small-angle deflections we
must adopt a different approach.  Consider again our star moving
through a field of similar stars.  For a single encounter with impact
parameter $b$, the resulting velocity change transverse to the
incoming velocity $v$ may be shown to be (see, e.g., \cite{2008gady.book.....B})
\begin{equation}
    \delta v_\perp = 2v\left(\frac{b}{b_{90}}\right)
			\left(1+\frac{b^2}{b_{90}^2}\right)^{-1}\,.
\end{equation}
Integrating over repeated random encounters, we expect the mean
velocity change in any direction transverse to the incoming velocity
to be zero, by symmetry.  However, the transverse velocity undergoes a
symmetric, two-dimensional random walk, and we expect transverse
velocity changes to add in quadrature, leading to a non-zero value for
the mean square transverse velocity $\Delta v_\perp^2$.  During a time
interval $\delta t$, the number of encounters with impact parameters
in the range $[b,b+db)$ is $2\pi b \,db \,nv\delta t$, so integrating
over all encounters, we find
\begin{eqnarray}
    \Delta v_\perp^2 &=& 2\pi nv\delta t\,
	\int_0^{b_{max}}\,b \,db \, \left(\delta v_\perp\right)^2 \nonumber\\
        &\approx& 8\pi\delta t\,\frac{G^2m_\ast^2n}{v}\,
		\ln\left(\frac{b_{max}}{b_{90}}\right)\,,
\end{eqnarray}
where we have assumed $b_{max}\sim\rvir\gg b_{90}$.  

We can define a two-body relaxation timescale, $\delta t_r$, as the
time interval in the above expression corresponding to $\Delta
v_\perp^2 = v^2$.  Rearranging the equation and replacing all
quantities by mean values, as above, we find
\begin{equation}
    \delta t_r ~=~ \frac{\vdisp^{3/2}}{8\pi G^2\mbar\rho\,\ln\Lambda}\,,
\label{McMillan:Eq:tr_basic}
\end{equation}
where the ``Coulomb logarithm''\index{Coulomb logarithm term} term (the term stemming from the
almost identical development found in plasma physics) has $\Lambda =
\rvir/b_{90} = \half N$, from Eqs.~(\ref{McMillan:Eq:vdisp_rvir}) and
(\ref{McMillan:Eq:b90}).

There is considerable ambiguity in the above definition.  For example,
we could equally well have used $\Delta v_\parallel^2$ as our measure
of relaxation, and our procedure neglects the distribution of relative
velocities of stars in a real system.  In fact, all approaches and
refinements yield the same functional dependence on physical
parameters as Eq.~(\ref{McMillan:Eq:tr_basic}), but they differ in the
numerical coefficient.  The expression presented in \cite{1987degc.book.....S},
now widely adopted as a standard definition of the term, defines the
relaxation time in terms of $\Delta v_\parallel^2$, and averages over
a thermal velocity distribution --- the theoretical end point (not
always realised in practice) of the relaxation process.  The result is
\begin{eqnarray}
    t_r &=& \frac{0.065\,\vdisp^{3/2}}{G^2\mbar\rho\,\ln\Lambda}
    \label{McMillan:Eq:tr}\\
	 &=& 3.4~\mbox{Gyr}
		 \left(\frac{\vdisp^{1/2}}{10\,\rm km/s}\right)^3
		 \left(\frac{\mbar}{\msun}\right)^{-1}
                 \left(\frac{\rho}{100\,\msun {\rm pc}^{-3}}\right)^{-1}
		 \left(\frac{\ln\Lambda}{10}\right)^{-1}\,.\nonumber
\end{eqnarray}

The precise definition of $\Lambda$ is also the subject of a minor
debate.  Spitzer \cite{1987degc.book.....S} chooses $b_{max}=\rh$ and hence writes
$\Lambda=0.4N$.  Giersz \& Heggie \cite{Giersz1994} calibrate the relaxation
process using $N$-body simulations (see
\S\ref{McMillan:Subsec:directnbody}) and find $\Lambda\sim0.1N$.  For
systems with a significant range of stellar masses, the effective
value of $\Lambda$ may be considerably smaller even than this value.

Although the analysis leading to Eq.~(\ref{McMillan:Eq:tr}) is global
in nature, it is common to find this expression used as a {\em local}
measure of the relaxation timescale in a system.

\subsection{Comparison of Timescales}
\label{McMillan:Subsec:comparison}

Comparing Eqs.~(\ref{McMillan:Eq:tr}) and (\ref{McMillan:Eq:ts}), we
see that
$$
    \frac{t_s}{t_r} ~\sim~ 5\ln\Lambda ~\sim~ 60 ~~\mbox{for}~~ N \sim 10^6,
$$
so distant encounters dominate over close encounters in determining
the flow of energy around the system.  

Spitzer \cite{1987degc.book.....S} defines a global relaxation\index{global relaxation} timescale, often referred
to as the {\em half-mass relaxation time}\index{half-mass relaxation time}, $\trh$, by replacing all
quantities in Eq.~(\ref{McMillan:Eq:tr}) with their system-wide
averages,
\begin{eqnarray*}
    \vdisp &\rightarrow& \frac{GM}{2\rvir}\\
    \rho~~ &\rightarrow& \frac{3M}{8\pi\rh^3}\\
    \mbar  &\rightarrow& \frac{M}{N},
\end{eqnarray*}
obtaining
\begin{eqnarray}
    \trh &=& \frac{0.138\,N\rh^{3/2}}{G^{1/2}M^{1/2}\ln\Lambda}
	\label{McMillan:Eq:trh}\\
\noalign{\smallskip}
	  &=& 6.5~\mbox{Gyr}
		 \left(\frac{N}{10^6}\right)
		 \left(\frac{M}{10^6\msun}\right)^{-1/2}
                 \left(\frac{\rh}{10~{\rm pc}}\right)^{3/2}
		 \left(\frac{\ln\Lambda}{10}\right)^{-1}\,.\nonumber
\end{eqnarray}
Hence, from Eqs.~(\ref{McMillan:Eq:tdyn}) and (\ref{McMillan:Eq:trh}),
we have
\begin{equation}
    \frac{\trh}{\tdyn} ~\sim~ \frac{N}{5\ln\Lambda}\,,
\label{McMillan:Eq:trtd}
\end{equation}
and we see that relaxation is a slow process relative to the dynamical
time for all but the smallest systems.

We note in passing that the relaxation time (Eq.~\ref{McMillan:Eq:tr})
evaluated at the half-mass radius $\rh$ can differ significantly from
the half-mass relaxation time (Eq.~\ref{McMillan:Eq:trh}) --- for
example, for a $W_0=7$ King model, the former exceeds the latter by a
factor of almost five --- representing a potentially significant source of
confusion in this terminology.

\subsection{Cluster Dynamical Evolution}
\label{McMillan:Subsec:evolution}

We can understand most aspects of the dynamical evolution of globular
clusters in terms of the fundamental physics of self-gravitating
systems just described.

\subsubsection{Evaporation and Cluster Lifetimes}
\label{McMillan:Subsubsec:evaporation}

The relaxation time (Eq.~\ref{McMillan:Eq:tr}) is the timescale on
which stars tend to establish a Maxwellian velocity distribution\index{Maxwellian velocity distribution}.  A
fraction $\xi_e$ of the stars in the tail of that distribution have
velocities larger than $\vesc$ and therefore escape.  If this
high-velocity tail is refilled every $\trh$, then the dissolution time
scale is $\tdis\sim\trh/\xi_e$.  For isolated clusters,
$\vesc=2\,\vrms$ and $\xi_e=0.0074$, implying $\tdis=137\,\trh$.  For
tidally limited clusters, $\xi_e$ is higher since $\vesc$ is lower.
For a ``typical'' cluster density profile (with $\rh/RJ=0.145$),
Spitzer  \cite{1987degc.book.....S} finds $\xi_e\approx0.045$, so $\tdis\approx20\,\trh$.

In fact, $\tdyn$ also enters into the escape rate\index{escape rate}.  

Baumgardt \cite{2001MNRAS.325.1323B} found, for equal-mass stars,
$\tdis\propto\trh^{3/4}\tdyn^{1/4}$.  This non-intuitive scaling of
the dissolution time results from the fact that a star with sufficient
energy to escape may orbit the system many times before finding one of
the Lagrangian points\index{Lagrangian point}, through which it eventually escapes
\cite{2000MNRAS.318..753F}.

Baumgardt \& Makino \cite{2003MNRAS.340..227B} found that this non-linear scaling of the
dissolution time with the relaxation time also holds for model
clusters with a stellar mass spectrum, stellar evolution, and for
different types of orbits in a logarithmic potential.  Their result
for $\tdis$ may be summarised as
\begin{equation}
\tdis\approx2\,{\rm Myr} 
            \left(\frac{N}{\ln\Lambda}\right)^{3/4}
            \left(\frac{R_G}{{\rm kpc}}\right)
            \left(\frac{V_G}{220\,{\rm km/s}}\right)^{-1}(1-\varepsilon),
\label{McMillan:Eq:tdis}
\end{equation}
where $\varepsilon$ is the ellipticity of the orbit.  For non-circular
orbits ($\varepsilon > 0$), the galactocentric distance $R_G$ is taken
as apogalacticon, while $V_G$ is the circular velocity, which is
constant in a logarithmic potential.  
Lamers, Gieles, \& Portegies Zwart \cite{2005A&A...429..173L} found that, when the Coulomb
logarithm is taken into account, the scaling is approximately
$\tdis\propto N^{0.65}$ for $M\sim10^3-10^6\,\msun$.

The combined effects of mass loss by stellar evolution and dynamical
evolution in the tidal field of the host galaxy have been extensively
studied by a number of authors, including
\cite{1990ApJ...351..121C,1995MNRAS.276..206F,2000ApJ...535..759T,2003MNRAS.340..227B,Whitehead2013}.
Mass loss\index{mass loss} due to stellar evolution, particularly during a
cluster's early evolution (the first few hundred million years), can
significantly reduce the cluster lifetime.  Cluster expansion due to
this mass loss can be substantial, and may even result in complete
disruption if the cluster is mass segregated\index{mass segregation} before the bulk of the
stellar evolution takes place
\cite{2009ApJ...698..615V}.

The expansion of a mass-segregated cluster will not be homologous\index{homologous expansion}, as
the massive (segregated) core stellar population tends to lose
relatively more mass than the lower-mass halo stars. The result is a
more dramatic expansion of the cluster core, with less severe effects
farther out.  These above studies show that when clusters expand to a
half-mass radius of $\sim0.5\,\rj$ they lose equilibrium and most of
their stars overflow $\rj$ in a few crossing times.

\subsubsection{Core Collapse}\index{core collapse}
\label{McMillan:Subsubsec:corecollapse}

The evaporation of high-velocity stars and the internal effects of
two-body relaxation transfer energy from the inner to the outer
regions of the cluster, resulting in the phenomenon of core collapse
\cite{1962spss.book.....A,1968MNRAS.138..495L,1980ApJ...242..765C,1980MNRAS.191..483L,Makino1996}.
During this phase, the central portions of the
cluster accelerate toward infinite density while the outer regions
expand.  The process is most easily understood by recognising that,
according to the virial theorem (Eq.~\ref{McMillan:Eq:specificheat}),
a self-gravitating system has negative specific heat\index{specific heat} --- reducing its
energy causes it to heat up.  Hence, as relaxation transports energy
from the (dynamically) warmer central core to the cooler outer
regions, the core contracts and heats up as it loses energy.  The time
scale for the process to go to completion (i.e. a core of zero size
and formally infinite density) is $\tcc\sim 15 \trh$ for an initial
Plummer sphere\index{Plummer sphere} of identical masses.  Starting with a more concentrated
King\index{King model} \cite{1966AJ.....71...64K} distribution shortens the time of core collapse
considerably
\cite{1996NewA....1..255Q},
as does a broad spectrum of masses
\cite{1985ApJ...292..339I}.

In systems with a mass spectrum, the dynamical evolution is
accelerated by the tendency of the system to evolve toward energy
equipartition\index{energy
equipartition}, in which the velocity dispersions of stars of different
masses would have $\langle mv^2\rangle\sim {\rm constant}$.  The
result is mass segregation\index{mass segregation}, where more massive stars slow down and
sink toward the centre of the cluster on a timescale
\cite{1969ApJ...158L.139S}
\begin{equation}
	\tseg \sim \frac{\langle m\rangle}{m}\,\trh \,.
\label{McMillan:Eq:tseg}
\end{equation}
Portegies Zwart \& McMillan \cite{2002ApJ...576..899P} and G\"urkan, Freitag, \& Rasio
\cite{2004ApJ...604..632G} find that, for a typical 
Kroupa \cite{2001MNRAS.322..231K} initial mass function, the timescale for the most massive stars
to reach the centre and form a well defined high-density core is
$\sim0.2 \trl$, where $\trl$ is the relaxation time
(Eq.~\ref{McMillan:Eq:tr}) of the region containing a significant
number of massive stars --- the core of a massive cluster, or the
half-mass radius of a smaller one (in which case $\trl = \trh$, see
Eq.~\ref{McMillan:Eq:trh}).

The post-collapse phase may involve a series of large-amplitude core
oscillations, driven by the same basic instability as gravothermal
collapse\index{gravothermal
collapse} and involving the innermost few percent of the mass.  First
discovered in gas-sphere \cite{1984MNRAS.208..493B,Goodman1987}
and later in Fokker--Planck \cite{Cohn1989} simulations (see
\S\ref{McMillan:Sec:modeling}), their existence in simple $N$-body
systems was subsequently confirmed by Makino \cite{Makino1996}.  Like core
collapse, these gravothermal oscillations\index{gravothermal oscillation} appear to be a ubiquitous
phenomenon.  However, they are known to be suppressed by the
presence of mass spectrum, as well as by primordial binaries and other
heating mechanisms, and their reality in actual globular clusters
remains unclear.  However, significant core oscillations have been
observed in realistic simulations of globular-cluster sized systems
(Heggie 2012, private communication), and the effect of these
substantial variations in central density may be important for the
formation and subsequent evolution of exotica such as neutron star\index{neutron star}
binaries and blue stragglers \cite{Grindlay2006}.

\subsection{Internal Heating}
\label{McMillan:Subsubsec:heating}

On longer timescales, cluster evolution is driven by the competition
between relaxation and a variety of internal heating mechanisms.  High
core densities lead to interactions among stars and binaries.  Many of
these interactions can act as energy sources to the cluster on larger
scales, satisfying the relaxation-driven demands of the halo and
temporarily stabilising the core against collapse
\cite{1989Natur.339...40G,1991ApJ...370..567G,1990ApJ...362..522M,1991ApJ...372..111M,
 1992MNRAS.257..513H,2003ApJ...593..772F}.
On long time
scales, these processes lead to a relatively slow (relaxation time
scale) overall expansion of the cluster, with $\rvir\propto t^{2/3}$,
a result that follows from simple considerations of the energy flux
through the half-mass radius
\cite{1965AnAp...28...62H}.

\subsubsection{Binary Interactions}
\label{McMillan:Subsubsec:binaries}

Binaries in star clusters may be primordial (i.e. they were present
when the cluster formed, or are descended from such systems), or they
can form in a variety of ways, including dissipationless
stellar-dynamical processes 
\cite{1987degc.book.....S, Tanikawa2012}
and
dissipative processes such as tidal capture \cite{Fabian1975}.
Regardless of how they formed, binaries are described by dynamicists
as either ``hard'' or ``soft,'' \index{hard binary}\index{soft binary}depending on their binding energies.
A hard binary has binding energy greater than the mean stellar kinetic
energy in the cluster
\cite{1975MNRAS.173..729H}:
$|E_b| > \frac12\langle mv^2\rangle \approx
\frac12\mbar\vdisp$, where $\mbar$ and $\vdisp$ are the local mean
stellar mass and velocity dispersion.  A binary with mass
$m_b=M_1+M_2$ and semi-major axis $a_b$ has energy $E_b =
-GM_1M_2/2a_b$, so hard binaries have $a_b<a_{\rm hard}$, where
\begin{equation}
	a_{\rm hard} ~=~ {G m_b^2 \over 4 \mbar \vdisp} ~\approx~ 
			    9.5 \times 10^{4} \,\rsun 
                            \left(\frac{m_b}{\msun}\right)^2  
                            \left(\frac{\vrms}{\rm km/s}\right)^{-2}.
\label{McMillan:Eq:hard}
\end{equation}
Here we have assumed that $M_1=M_2=\mbar$ in deriving the right-hand
expression.  The hard--soft distinction is often helpful when
discussing dynamical interactions between binaries and other cluster
members.  However, since this definition of hardness depends on local
cluster properties, the nomenclature changes with environment --- a
binary that is hard in the halo could be soft in the core.

The dynamical significance of hard binaries (see
Eq.~\ref{McMillan:Eq:hard}) has been understood since the 1970s
\cite{1975MNRAS.173..729H,1975AJ.....80..809H,1983ApJ...268..319H}.
When a hard binary
interacts with another cluster star, the resultant binary (which may
or may not have the same components as the original binary) tends, on
average, to be harder than the original binary, making binary
interactions a net heat source to the cluster.  Soft binaries tend to
be destroyed by encounters.  For equal-mass systems, the mean energy
liberated in a hard-binary encounter is proportional to $E_b$:
$\langle\Delta E_b\rangle = \gamma E_b$, where $\gamma = 0.4$ for
``resonant'' interactions 
\cite{1975MNRAS.173..729H},
and $\gamma\sim0.2$ when wider ``flybys'' are taken
into account
\cite{1987degc.book.....S}.

The liberated energy goes into the recoil of the binary and single
star after the interaction.  Writing the binary energy as $E_b=-hkT$
(see \S\ref{McMillan:Subsec:scales}), where $h\gg1$, the total recoil
energy, in the centre of mass frame of the interaction, is $\gamma
hkT$.  A fraction ${m_b \over m_b + m}$ of this energy goes to the
single star (of mass $m$) and ${m \over m_b + m}$ to the binary.  For
equal-mass stars, these fractions reduce to $\frac23$ for the single
star and $\frac13$ for the binary.  Neglecting the thermal motion of
the centre of mass frame, we can identify three regimes:
\begin{enumerate}
\item If $\frac23\gamma hkT < \frac12m\vesc^2 = 2m\vdisp = 6kT$,
  i.e. $h<45$, neither the binary nor the single star acquires enough
  energy to escape the cluster. Binaries in this stage are kicked out
  of the core, then sink back by dynamical friction\index{dynamical friction}.
\item If $\frac23\gamma hkT > 6kT$ but $\frac13\gamma hkT < 4m\vdisp
  = 12kT$, i.e. $45<h<180$, the single star escapes, but the binary is
  retained.
\item If $h > 36/\gamma = 180$, both the binary and the single star
  are ejected.
\end{enumerate}
These numbers are valid only for equal-mass stars, and are intended
for illustration only.  For a binary with components more massive than
average, as is often the case, the threshold for single star ejection
drops, while that for self-ejection increases.

The binary encounter timescale is $\tenc = (n \sigma \vrms)^{-1}$,
where $n$ is the local stellar density and $\sigma$ is the encounter
cross section (see Eq.~\ref{McMillan:Eq:crosssection}).  If we
arbitrarily compute the binary interaction cross section as that for a
flyby within 3 binary semi-major axes, consistent with the encounters
contributing to the
Spitzer \cite{1987degc.book.....S} value $\gamma=0.2$, and again assume equal masses ($m_b
= 2m$), we find
\begin{equation}
	\tenc \sim 8h\tr ,
\end{equation}
where we have used Eq.~(\ref{McMillan:Eq:tr}) and taken $\ln\Lambda=10$.
Thus the net local heating rate per binary during the 100\% efficient
phase (\#1 above), when the recoil energy\index{recoil energy} remains in the cluster, is
\begin{equation}
    \gamma hkT\,\tenc^{-1} \sim 0.1kT/\tr,
\end{equation}
that is, on average, each binary heats the cluster at a roughly
constant rate.  During phase 2, the heating rate drops to just over
one-third of this value.  The limiting value of one-third is not
reached since the ejected single stars still heat the cluster
indirectly by reducing its binding energy by a few $kT$.  For
``self-ejecting'' binaries, the heating rate drops almost to zero,
with only indirect heating contributing.

Binary--binary interactions also heat the cluster, although the extra
degrees of freedom complicate somewhat the above discussion.  If the
binaries differ widely in semi-major axes, the interaction can be
handled in the three-body approximation, with the harder binary
considered a point mass.  If the semi-major axes are more comparable,
as a rule of thumb the harder binary tends to disrupt the wider one
\cite{1996MNRAS.281..830B}.

Numerical experiments over the past three decades have unambiguously
shown how initial binaries segregate to the cluster core, interact,
and support the core against further collapse
\cite{1990ApJ...362..522M,1992MNRAS.257..513H}.
The respite
is only temporary, however.  Sufficiently hard binaries are ejected
from the cluster by the recoil from their last interaction, and
binaries may be destroyed by interactions with harder binaries, or by
collisions during the interaction.  For large initial binary
fractions, this binary-supported phase may exceed the age of the
universe or the lifetime of the cluster against tidal dissolution\index{tidal dissolution}.
However, for low initial binary fractions, as appears to have been the
case for the globular clusters\index{globular cluster} observed today
\cite{2008MmSAI..79..623M},
 the binaries can be depleted before the cluster
dissolves, and core collapse resumes
\cite{2003ApJ...593..772F}.

\subsubsection{Stellar Collisions}
\label{McMillan:Subsubsec::collisions}

In systems without significant binary fractions --- either initially or
following the depletion of core binaries --- core collapse may continue
to densities at which actual stellar collisions\index{stellar collision} occur.  In young
clusters, the density increase may be enhanced by rapid segregation\index{mass segregation} of
the most massive stars in the system to the cluster core.  Since the
escape velocity from the stellar surface greatly exceeds the \emph{rms} speed
of cluster stars, collisions are expected to lead to mergers\index{merger} of the
stars involved, with only small fractional mass loss
\cite{1987ApJ...323..614B,2001A&A...375..711F}.
If the merger products did
not evolve, the effect of collisions\index{collision} would be to dissipate kinetic
energy, and hence cool the system, accelerating core collapse
\cite{1999A&A...348..117P}. 
However, when accelerated stellar
evolution is taken into account, the (time averaged) enhanced mass
loss can result in a net heating effect
\cite{2008IAUS..246..151C}.

The cross section for an encounter between two objects of masses $M_1$
and $M_2$ and radii $r_1$ and $r_2$, respectively, is \cite{1976ApL....17...87H}
\begin{equation}
	\sigma = \pi r^2 \left[ 1 + {2Gm \over r v^2} \right] ,
\label{McMillan:Eq:crosssection}
\end{equation}
where $v$ is the relative velocity at infinity,
$m = M_1+M_2$, and $r = r_1+r_2$.  For $r \ll Gm/v^2$, as is usually
the case for the objects of interest here, the encounter is dominated
by the second term (gravitational focusing), and
Eq.~(\ref{McMillan:Eq:crosssection}) reduces to
\begin{equation}
	\sigma \approx 2 \pi r {G m \over v^2}.
\label{McMillan:Eq:crosssection2}
\end{equation}

Collisions between unbound single stars are unlikely unless one or
both of the stars is very large and/or very massive, or the local
density is very high.  However, the presence of a substantial binary
population can significantly increase the chance of a stellar
collision.  The closest approach between particles in a resonant
interaction may be as little as a few percent of the binary semi-major
axis
\cite{1985ApJ...298..502H},
so the hardest binaries may well experience
physical stellar collisions rather than hardening to the point of
ejection.  It is quite likely that the third star will also be
engulfed in the collision product
\cite{2004MNRAS.352....1F}.
Alternatively, before its next interaction,
the binary may enter the regime in which internal processes, such as
tidal circularisation\index{tidal circularisation} and/or Roche lobe overflow\index{Roche lobe overflow}, become important.
The future of such a binary may be determined by the internal
evolution of its component stars, rather than by further encounters.

Since binaries generally have semi-major axes much greater than the
radii of the component stars, these binary-mediated collisions play
important roles in determining the stellar collision rate in most
clusters
\cite{2002ApJ...576..899P},
leading to significant numbers of
mergers in lower-density, binary rich environments.  Massive binaries
in dense clusters tend to be collision targets rather than heat
sources
\cite{2004ApJ...604..632G}.

\section{Multiple Stellar Populations}
\label{McMillan:Sec:multiplepop}

The discovery of multiple populations\index{multiple populations} of main-sequence stars and
giants in an increasing number of globular clusters
\cite{2005ApJ...621..777P,2008MmSAI..79..334P}
has led to the realisation that
these clusters are not idealised entities with single well defined
stellar populations.  In many systems the observed stellar populations
appear to be separated by less than $\sim 10^8$ years.  The existence
of multiple populations suggests that a second epoch of star formation
must have taken place early in the cluster's lifetime.  The
differences in light-element abundances suggest that the
second-generation (SG) stars formed out of gas containing matter
processed through high-temperature CNO cycle\index{CNO cycle} reactions in
first-generation (FG) stars \cite{Carretta2009a,Carretta2009b}.

The origin of the gas from which SG stars form is still an open
question.  Current leading models involve AGB stars\index{AGB star}
\cite{2001ApJ...550L..65V, 2007PASA...24..103K},
rapidly rotating massive stars\index{massive star}
\cite{2006A&A...458..135P, 2007A&A...464.1029D},
and massive
binaries (\cite{deMink2009}, see also \cite{Renzini2008}
for a review).  In order to explain the observed abundance patterns,
all current models require that ``pristine'' (i.e.~unprocessed) gas
must be included in the SG mix (see \cite{Ercole2010} and
references therein).  In addition, in order to form the numbers of SG
stars observed today, the FG cluster must have been considerably more
massive than it is now, and the majority of stars in the cluster
initially belonged to the FG population.

Many fundamental questions concerning globular cluster star formation
and cluster chemical and dynamical history are raised by the discovery
of multiple populations, and they have been targets of numerous
theoretical investigations 
\cite{Ercole2008,Ercole2010,Bekki2011,Vesperini2011}.
Recently, Bastian et al. \cite{Bastian2013} have
described a scenario that avoids both the mass problem and the need
for multiple star-formation episodes by considering the accretion of
CNO enriched material onto still-forming protostellar discs.

In many cases, the models suggest that the SG (``enriched'')
population should initially be significantly more centrally
concentrated than the FG stars.  Decressin et al. \cite{Decressin2008} and Vesperini
et al. \cite{Vesperini2013} have studied the subsequent evolution and mixing of the
two-component cluster in the first scenario.  Vesperini et al. \cite{Vesperini2013}
find that the timescale for complete mixing depends on the SG initial
concentration, but in all cases complete mixing is expected only for
clusters in advanced evolutionary phases, having lost lost at least
60--70 percent of their mass due to two-body relaxation\index{two-body relaxation}.  These
scenarios may be relevant to the properties of blue stragglers because
they suggest that the FG and SG binary populations should have
significantly different dynamical histories, with the SG binaries
having spent much of their lives in much denser environments.  One
might naively expect these differences to manifest themselves in the
properties of FG and SG blue stragglers, although the limited data
currently available give little hint of any such effect (see Chap. 5).

\section{Modeling Star Clusters}
\label{McMillan:Sec:modeling}

Although the fundamental physics is not hard to understand, simulating
star clusters\index{star cluster} can be a complex numerical undertaking.  Significant
complications arise due to the long-range nature of the gravitational
force, which means that every star in the cluster is effectively in
constant communication with every other, leading to high computational
cost.  Further complications arise from the enormous range in spatial
and temporal scales inherent in a star cluster.  Computers, by the way
they are constructed, have difficulty in resolving such wide ranges,
and many of the software problems in simulations of self-gravitating
systems\index{self-gravitating
system} arise from this basic limitation.  The combination of many
physical processes occurring on many scales, with high raw processing
requirements, makes numerical gravitational dynamics\index{gravitational dynamics} among the most
demanding and challenging areas of computational science.  Here we
discuss some issues involved in the numerical modeling of massive star
clusters.

A broad spectrum of numerical methodologies is available for
simulating the dynamical evolution of globular clusters.  In
approximate order of increasing algorithmic and physical complexity,
but not necessarily in increasing numerical complexity, the various
methods may be summarised as follows.

\begin{itemize}
\item {\em Static Models} \index{static model} are self-consistent potential--density pairs
  for specific choices of phase-space distribution functions
  \cite{1911MNRAS..71..460P,1966AJ.....71...64K,2008gady.book.....B}.
  They
  have been instrumental in furthering our understanding of cluster
  structure, and provide a framework for semi-analytical treatments of
  cluster dynamics.  However, they do not lend themselves to detailed
  study of star cluster evolution, and we will not discuss them
  further here.

\item {\em ``Continuum'' Models}\index{continuum model} treat the cluster as a quasi-static
  continuous fluid whose phase-space distribution function evolves
  under the influence of two-body relaxation and other energy sources
  (such as binary heating) that operate on relaxation timescales
  (Eq.~\ref{McMillan:Eq:trh}).

\item {\em Monte Carlo Models} \index{Monte-Carlo method}treat some or all components of the
  cluster as pseudo-particles whose statistical properties represent
  the continuum properties of the system, and whose randomly chosen
  interactions model relaxation and other processes driving the
  long-term evolution.

\item {\em Direct $N$-body Models} \index{N-body model}follow the individual orbits of all
  stars in the system, automatically including dynamical and
  relaxation processes, and modeling other physical processes on a
  star-by-star basis.
\end{itemize}

Much of our current understanding of the evolution of star clusters
comes from detailed numerical simulations\index{numerical simulation}, and the above techniques
are used for the vast majority of simulations.  Here we present a few
details of these simulation techniques. We end with brief discussions
of new computer hardware and the state of the art in modern simulation
codes.

\subsection{Continuum Methods}
\label{McMillan:Subsec:continuum}

The two leading classes of continuum models are gas-sphere
\cite{1980MNRAS.191..483L,1984MNRAS.208..493B,2001A&AT...20...47D}
and Fokker--Planck\index{Fokker--Planck method}
\cite{1979ApJ...234.1036C,1985IAUS..113..373S,1990ApJ...351..121C,1992ApJ...386..106D,1996PASJ...48..691T,1997PASJ...49..547T,1998ApJ...503L..49T}
methods.  They have mainly been applied to spherically symmetric
systems, although axisymmetric extensions to rotating systems have
also been implemented
\cite{1999MNRAS.302...81E,2002MNRAS.334..310K,2004MNRAS.351..220K},
and a few limited experiments with simplified binary treatments have
also been carried out
\cite{1991ApJ...370..567G}.

Both approaches start from the collisional Boltzmann equation\index{Boltzmann equation} as the
basic description for a stellar system, then simplify it by averaging
the distribution function $f({\bf x}, {\bf v})$ in different ways.
Gas-sphere methods proceed in a manner analogous to the derivation of
the equations of fluid motion, taking velocity averages to construct
the moments of the distribution: $\rho=\int\,d^3v\,f({\bf x}, {\bf
  v})$, ${\bf u}=\int\,d^3v\,{\bf v}f({\bf x}, {\bf v})$, $\sigma^2=
\frac13 \int\,d^3v\,v^2f({\bf x}, {\bf v})$, etc.  Application of an
appropriate closure condition leads to a set of equations identical to
those of a classical conducting fluid, in which the conductivity
depends inversely on the local relaxation time.  Fokker--Planck
methods transform the Boltzmann equation by orbit-averaging all
quantities and recasting the equation as a diffusion equation in $E-J$
space, where $E$ is stellar energy and $J$ is angular momentum.  Since
both $E$ and $J$ are conserved orbital quantities in a static,
spherically symmetric system, two-body relaxation enters into the
problem via the diffusion coefficients.

These methods have been of enormous value in developing and refining
theoretical insights into the fundamental physical processes driving
the dynamical evolution of stellar systems
\cite{1984MNRAS.208..493B}.
However, as the degree of realism
demanded of the simulation increases --- adding a mass spectrum, stellar
evolution, binaries, etc. --- the algorithms rapidly become cumbersome,
inefficient, and of questionable validity
\cite{1999CeMDA..73..179P}.

\subsection{Monte Carlo Methods}\index{Monte-Carlo method}
\label{McMillan:Subsec:montecarlo}

Depending on one's point of view, Monte Carlo methods can be regarded
as particle algorithms for solving the partial differential equations
arising from the continuum models, or approximate schemes for
determining the long-term average gravitational interactions of a
large collection of particles.  The early techniques developed in the
1970s and 1980s 
\cite{1971ApJ...164..399S,1973dses.conf..183H,1975IAUS...69....3S,1982AcA....32...63S,1986AcA....36...19S}
fall into the former category, but more recent studies
\cite{1998MNRAS.298.1239G,2000ApJ...540..969J,2001A&A...375..711F,2001MNRAS.324..218G,2003ApJ...593..772F,2006MNRAS.371..484G,2007ApJ...658.1047F,2008MNRAS.389.1858H,2009MNRAS.395.1173G,Giersz2011,Giersz2013},
tend
to adopt the latter view.  The hybrid Monte Carlo scheme of
\cite{1998MNRAS.298.1239G,2001MNRAS.324..218G,2003MNRAS.343..781G}
combines a gas-sphere
treatment of the ``background'' stellar population with a Monte Carlo
realisation of the orbits and interactions of binaries and other
objects of interest.  These approaches allowed the first simulations
of an entire globular cluster, from a very early (although gas
depleted) phase to complete dissolution.

Monte Carlo methods are designed for efficient computation of
relaxation effects in collisional stellar systems, a task which they
accomplish by reducing stellar orbits to their orbital
elements --- energy and angular momentum --- effectively orbit averaging
the motion of each star.  Relaxation is modeled by randomly selecting
pairs of stars and applying interactions between them in such a way
that, on average, the correct rate is obtained.  This may be
implemented in a number of ways, but interactions are generally
realised on timescales comparable to the orbit-averaged relaxation
time.  As a result, Monte Carlo schemes can be orders of magnitude
faster than direct $N$-body codes\index{N-body code}.  To achieve these speeds, however,
the geometry of the system must be simple enough that the orbital
integrals can be computed from a star's instantaneous energy and
angular momentum\index{angular momentum}.  In practice, this limits the approach to
spherically symmetric systems in virial equilibrium\index{virial equilibrium}, and global
dynamical processes occurring on relaxation\index{relaxation} (or longer) timescales.

\subsection{$N$-body Methods}
\label{McMillan:Subsec:directnbody}

$N$-body codes incorporate detailed descriptions of stellar dynamics
at all levels, using direct integration of the individual (Newtonian)
stellar equations of motion for all stars
\cite{2003gnbs.book.....A,2003gmbp.book.....H}.
Their major attraction is that
they are assumption-free, in the sense that all stellar interactions
are automatically included to all orders, without the need for any
simplifying approximations or the inclusion of additional reaction
rates to model particular physical processes of interest.  Thus,
problems inherent to Fokker--Planck and Monte Carlo methods related to
departures from virial equilibrium, spherical symmetry, statistical
fluctuations, the form of (and indeed the existence of) phase space
distribution functions, and the possibility of interactions not
explicitly coded in advance, simply do not arise, and therefore do not
require fine-tuning as in the Monte Carlo models.

The price of these advantages is computational expense.  Each of the
$N$ particles must interact with every other particle a few hundred
times over the course of every orbit, each interaction requires $O(N)$
force calculations, and a typical (relaxation time) run spans $O(N)$
orbits (see Eq.~\ref{McMillan:Eq:trtd}).  The resulting $O(N^3)$
scaling of the total CPU time\index{CPU time} means that, even with the best time-step
algorithms, integrating even a fairly small system of a few hundred
thousand stars requires sustained teraflops speeds for several months
\cite{1988Natur.336...31H}.
Radically improved performance can be
achieved by writing better software, or by building faster computers
(or both).  However, the remarkable speed-up of $N$-body codes over
the last four decades has been due mainly to advances in hardware.

Substantial performance improvements were realised by adopting better
(individual) time stepping schemes\index{time stepping scheme} (as opposed to earlier shared time
step schemes), in which particles advance using steps appropriate to
their individual orbits, rather than a single step for all.  Block
time step schemes
\cite{1986LNP...267..156M,2006NewA...12..124M}
offer still better performance.
Further gains were made by utilising neighbour schemes
\cite{1973ApJ...179..885A},
which divide the force on every particle into
irregular (rapidly varying) and regular (slowly varying) parts, due
(loosely speaking) to nearby and more distant bodies.  By recomputing
the regular force at every particle step, but extrapolating the more
expensive $O(N)$ regular force for most time steps, and recomputing it
only on longer timescales, significant improvements in efficiency are
realised.  These schemes form the algorithmic basis for Aarseth's
{\tt NBODY6}\index{NBODY6 code} \cite{2003gnbs.book.....A} and its parallel counterpart {\tt NBODY6++}
\cite{1999JCoAM.109..407S}.

The 1980s saw a major algorithmic improvement with the development of
tree codes\index{tree code}
\cite{1986Natur.324..446B},
which reduce the force calculation complexity
from $O(N)$ to $O(\log N)$.  However, despite their algorithmic
efficiency, tree codes have not been widely used in modeling
collisional systems (but see 
\cite{1993ApJ...414..200M}).
This may be due to lingering
technical concerns about their long-term accuracy in systems dominated
by relaxation processes and their performance in clusters with large
dynamic ranges in densities and timescales, even though these
objections may not be well founded
\cite{1999MNRAS.310.1147M,2000ApJ...536L..39D}.

In recent years, meta-algorithms have been developed that enable
straightforward combination of previously distinct dynamical
algorithms to address new, more complex simulations.  The first
application of this approach permitted detailed study of the
interaction between a star cluster (modeled by a direct $N$-body) and
the surrounding galactic stellar population (modeled by a tree code)
\cite{2007PASJ...59.1095F,2009NewA...14..369P}.
This ``bridge''
scheme has subsequently been generalised \cite{Pelupessy2012,Pelupessy2013} to couple arbitrary dynamical
integration schemes, and now allows stellar- and gas-dynamical codes
to be combined in ways that were previously impossible to realise.

\subsection{Hardware Acceleration}
\label{McMillan:Subsec:acceleration}

The ``GRAPE''\index{GRAPE machine} (short for
``GRAvity PipE'') series of machines developed by Sugimoto and
co-workers at Tokyo University 
\cite{1993PASJ...45..269E}.
represented a quantum leap in gravitational
$N$-body simulation speed.  Abandoning algorithmic sophistication in
favour of simplicity and raw computing power, GRAPE systems achieved
high performance by mating a fourth-order Hermite integration scheme\index{Hermite integration scheme}
\cite{1992PASJ...44..141M}
with special-purpose hardware in the form of
highly parallel, pipelined accelerators implementing the computation\index{parallel computation}
of all inter-particle forces entirely in hardware.  Operationally, the
GRAPE hardware was simple to program, as it merely replaced the
function that computes the force on a particle by a call to hardware
interface libraries, leaving the remainder of the user's $N$-body code
unchanged.

In recent years, as in many specialty fields, the market appears to
have overtaken niche hardware solutions, and \emph{Graphics Processing Units}
(GPUs) have largely replaced GRAPEs in most $N$-body codes.  GPU\index{GPU}
accelerated codes now surpass the older GRAPE benchmarks for raw
performance and price/performance by a substantial margin.
Fortunately, the GRAPE-accelerated code has not been discarded, as
GPUs can serve as very efficient GRAPE emulators (see 
\cite{2007NewA...12..641P,2007astro.ph..3100H,2008NewA...13..103B,2009NewA...14..630G}
for various GPU implementations of
the GRAPE interface).  Besides GRAPE emulation, however, the much more
flexible programming model for GPUs (as well as the GRAPE-DR 
\cite{2005astro.ph..9278M}),
means that many other kinds of algorithms can (in principle) be
accelerated, although, in practice, it currently seems that
CPU-intensive operations such as direct $N$-body force summation show
substantially better acceleration than, say, tree codes running on the
same hardware.

Today, GPU-enabled code lies at the heart of almost all detailed
$N$-body simulations of star clusters and dense stellar systems.  The
GPU accelerated {\tt NBODY6-GPU}
 \cite{2012MNRAS.424..545N}
represents the current state of the art in
raw $N$-body speed on workstations, and numerous parallel,
GPU-accelerated $N$-body codes now exist or are under development,
including {\tt HiGPUs}\index{HiGPUs code} \cite{Capuzzo-Dolcetta2012},
	{ \tt phiGPU}\index{phiGPU code} \cite{Berczik2011},
{\tt ph4}\index{ph4 code} \cite{McMillan2011},
and {\tt NBODY6++} \cite{1999JCoAM.109..407S,Spurzem2008,Wang2013}
GPU accelerated versions of
sixth and eighth order extensions of the standard fourth-order
Hermite scheme 
\cite{2008NewA...13..498N}, 
with and without neighbour schemes, are also
becoming widespread.

\subsection{The Kitchen Sink}
\label{McMillan:Subsec:kitchensink}

The leading simulation programs in this field are ``kitchen sink''
packages that combine treatments of dynamics, stellar and binary
evolution, and stellar hydrodynamics within a single simulation.  Of
these, the most widely used are the $N$-body codes {\tt NBODY}\index{NBODY code}
\cite{2001MNRAS.323..630H,2003gnbs.book.....A},
(Hurley et al. 2001, Aarseth 2003), 
{\tt KIRA}\index{KIRA code} which is part of the
{\tt STARLAB}\index{STARLAB code} package
(e.g. \cite{2001MNRAS.321..199P}),
the {\tt MOCCA} Monte Carlo code\index{MOCCA code}
developed by Giersz and collaborators
\cite{1998MNRAS.298.1239G,2008MNRAS.389.1858H,2009MNRAS.395.1173G, Giersz2013},
and the Northwestern {\tt MC} Monte Carlo code\index{MC code}
\cite{2003ApJ...593..772F,2006MNRAS.368..121F,2007ApJ...658.1047F}.

Despite the differences in their handling of the large-scale dynamics,
these codes all employ similar approaches to stellar and binary
evolution\index{stellar evolution}\index{binary evolution} and collisions\index{collision}.  All use approximate descriptions of stellar
evolution, generally derived from look-up tables based on the detailed
evolutionary models of
\cite{1989ApJ...347..998E}
and
\cite{2000MNRAS.315..543H}.
They also rely on semi-analytic or
heuristic rule-based treatments of binary evolution
\cite{1996A&A...309..179P,2002MNRAS.329..897H},
conceptually similar from code to code, although significantly
different in detail and implementation.

In most cases, stellar collisions are implemented in the
``sticky-sphere'' approximation, where stars are taken to collide (and
merge) if they approach within the sum of their effective radii.  The
radii are calibrated using hydrodynamical simulations, and in some
cases mass loss is included in an approximate way.  Freitag's Monte
Carlo code, geared mainly to studies of galactic nuclei, interpolates
encounter outcomes from a pre-computed grid of smoothed Particle
Hydrodynamics (SPH) simulations\index{Smoothed Particle
Hydrodynamics}
\cite{2005MNRAS.358.1133F}.
Interesting alternatives, currently only operational in AMUSE\index{AMUSE code}
(see below), are the ``Make Me A Star'' package (MMAS; 
\cite{2003MNRAS.345..762L})
and its extension ``Make Me a Massive Star'' (MMAMS; 
\cite{2008MNRAS.383L...5G}).
They construct a merged
stellar model by sorting the fluid elements of the original stars by
entropy or density, then recomputing their equilibrium configuration,
using mass loss and shock heating data derived from SPH calculations.

Small-scale dynamics of multiple stellar encounters, such as binary
and higher-order encounters, are often handled by look-up from
pre-computed cross sections or --- more commonly --- by direct
integration, either in isolation or as part of a larger $N$-body
calculation.  Codes employing direct integration may also include
post-Newtonian terms in the interactions between compact objects
\cite{2006MNRAS.371L..45K}.

\subsection{The AMUSE Software Framework}
\label{McMillan:Subsec:future}

The comprehensiveness of kitchen-sink codes gives them the great
advantage of applicability to complex stellar systems, but also the
significant disadvantage of inflexibility.  By selecting such a code,
one chooses a particular hard-coded combination of dynamical
integrator, stellar and binary evolution schemes, collision
prescription, and treatment of multiple dynamics.  The structure of
these codes is such that implementing a different algorithm within the
larger framework is difficult at best for an expert, and impossible in
practice for others.

AMUSE\index{AMUSE code} (the \emph{Astrophysical Multipurpose Software Environment}) is a
collaborative effort begun in 2008 designed to address this class of
problem, providing a modular and extensible means of combining
individual ``monophysics'' solvers into a unified astrophysical
multiphysics simulation.  The overarching goal of the project is to
disentangle these components by providing a framework in which
individual modules can interoperate, to facilitate experimentation and
direct comparison of competing or alternative implementations of
specific physical processes.

\begin{figure}[h]
\includegraphics[width=119mm]{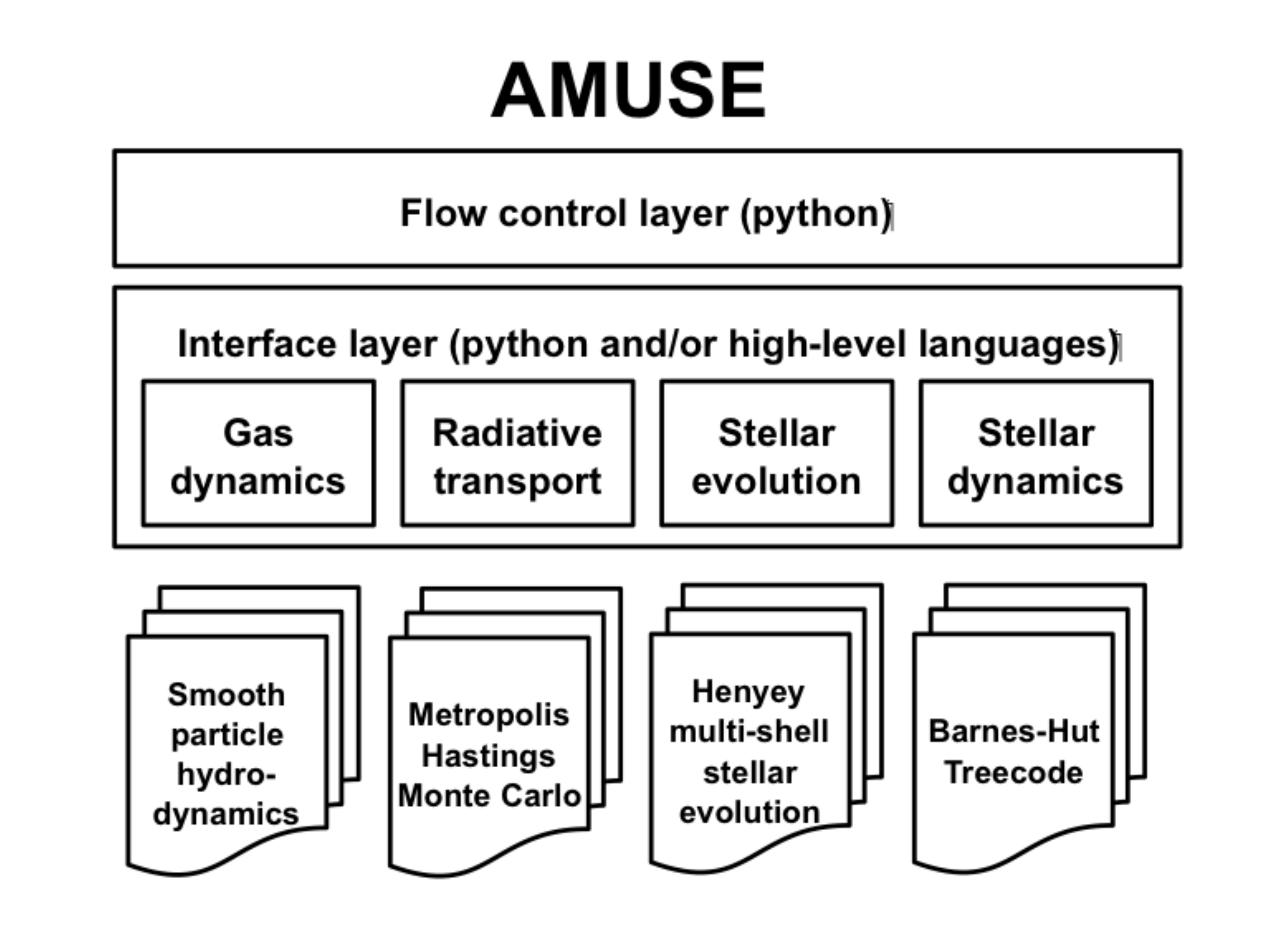}
\caption{The AMUSE environment.  The top-level flow control layer is
  typically a custom GUI or user-written Python script that specifies
  the structure of the program, effectively replacing the top-level
  loop of a traditional program.  Each of the four physics areas shown
  in the interface layer may be instantiated by one of several
  modules, allowing arbitrary combinations to be explored.}
\label{McMillan:Fig:AMUSE}
\end{figure}

The global structure of AMUSE is illustrated in
Fig.~\ref{McMillan:Fig:AMUSE}.  In the AMUSE programming model, each
piece of physics (advance the stellar or gas dynamics to a specified
time, manage a close encounter, evolve a star, collide two stars,
etc.) is implemented as a module with a standard interface onto the
rest of the system, but the model details are private to each module.
For example, all stellar modules include accessor functions that
provide information on the mass and radius of a specified star, but
the details of what a ``star'' actually is (an analytic formula, an
entry in a look-up table, or a set of 1- or 2-D arrays describing the
run of density, temperature, composition, etc.)  remain internal to
the module and are normally invisible to the outside.

The high-level ``glue'' language for AMUSE is python, chosen for its
rich feature set, ease of programming and rapid prototyping,
object-oriented capabilities, large user base in the astronomical
community, and extensive user-written software.  The design of AMUSE
places no restrictions on the choice of language for any given module.

In a typical application, the top-level loop (the flow control layer
in Fig.~\ref{McMillan:Fig:AMUSE}) of a simulation is written entirely
in python, allowing monitoring, analysis, graphics, grid management,
and other tools to be employed.  The relatively low speed of the
language does not significantly impact performance, because in
practice virtually all of the computational load is carried by the
(high-performance) physics modules.

Currently, AMUSE contains at least two (and typically more)
independent modules for each physical process supported, allowing
``plug and play'' interchangeability between implementations.  This
modular approach enables, for the first time in this area of
computational astrophysics, direct comparison and calibration of
different implementations of the same physical processes, and
facilitates experimentation in constructing new models.  The
integration of the parallel Message Passing Interface\footnote{http://www.mcs.anl.gov/mpi} (MPI) into AMUSE enables parallelism in
all modules, allowing a serial user script to manage and transparently
control modules that may themselves be parallel and/or GPU
accelerated, possibly running on remote high-performance clusters.

Details on the structure and applications of AMUSE may be found on
the project web site\footnote{http://amusecode.org} and in
\cite{Zwart2013,Pelupessy2013}.

\input{refmcmillan}